\documentclass[aps,prd,reprint,twocolumn,superscriptaddress,showpacs]{revtex4-1}
\usepackage{graphicx}
\usepackage{mathrsfs}
\usepackage{bm}
\usepackage{amsmath}
\usepackage{dcolumn}
\usepackage{epstopdf}
\usepackage{dsfont}
\usepackage{amssymb}
\usepackage{tabularx}
\usepackage{array}
\usepackage{float}
\usepackage{color}
\usepackage{epstopdf}
\usepackage{mathrsfs}
\usepackage[colorlinks, linkcolor=blue,anchorcolor=blue,citecolor=blue,urlcolor=blue]{hyperref}

\begin{document}
\title{Type-II topological metals}

\author{Si Li}
\email{sili@hunnu.edu.cn}
\affiliation{Key Laboratory of Low-Dimensional Quantum Structures and Quantum Control of Ministry of Education, Department of Physics and Synergetic Innovation Center for Quantum Effects and Applications, Hunan Normal University, Changsha 410081, China}
\affiliation{Research Laboratory for Quantum Materials, Singapore University of Technology and Design, Singapore 487372, Singapore}

\author{Zhi-Ming Yu}
\affiliation{Key Lab of Advanced Optoelectronic Quantum Architecture and Measurement (MOE),
Beijing Key Lab of Nanophotonics $\&$ Ultrafine Optoelectronic Systems, and School of Physics,
Beijing Institute of Technology, Beijing 100081, China}

\author{Yugui Yao}
\affiliation{Key Lab of Advanced Optoelectronic Quantum Architecture and Measurement (MOE),
Beijing Key Lab of Nanophotonics $\&$ Ultrafine Optoelectronic Systems, and School of Physics,
Beijing Institute of Technology, Beijing 100081, China}

\author{Shengyuan A. Yang}
\affiliation{Research Laboratory for Quantum Materials, Singapore University of Technology and Design, Singapore 487372, Singapore}

\begin{abstract}
Topological metals (TMs) are a kind of special metallic materials, which feature nontrivial band crossings near the Fermi energy, giving rise to peculiar quasiparticle excitations. TMs can be classified based on the characteristics of these band crossings. For example, according to the dimensionality of the crossing, TMs can be classified into nodal-point, nodal-line, and nodal-surface metals. Another important property is the type of dispersion.
According to degree of the tilt of the local dispersion around the crossing, we have type-I and type-II dispersions. This leads to significant distinctions in the physical properties of the materials, owing to their contrasting Fermi surface topologies.
In this article, we briefly review the recent advances in this research direction, focusing on the concepts, the physical properties, and the material realizations of the type-II nodal-point and nodal-line TMs.
\end{abstract}

\maketitle



\tableofcontents

\section{Introduction}
Topological metals (TMs) (including semimetals) have been attracting tremendous interest in current condensed-matter physics research~\cite{armitage2018weyl,burkov2016topological,chiu2016classification,yang2016dirac,bansil2016colloquium}. In TM materials, the electronic band structures possess nontrivial band crossings near the Fermi energy, such that the low-energy quasiparticles behave distinctly from the conventional Schr\"{o}dinger-type fermions, leading to unusual physical properties.

TMs are classified based on the characteristics of these band crossings. For example,
according to the dimensionality of the band crossing, which (for a 3D material) can be 0D, 1D, and 2D, we respectively have nodal-point, nodal-line, and nodal-surface TMs. TMs can also be classified by the degree of degeneracy of the band crossing. For example, a Weyl semimetal refers to a TM with twofold degenerate nodal points~\cite{murakami2007phase,wan2011topological}, whereas in a Dirac semimetal, the nodal points have fourfold degeneracy~\cite{young2012dirac,wang2012dirac,wang2013three}. Nodal features with other degrees of degeneracy, such as threefold, sixfold, or even eightfold, have also been discussed~\cite{bradlyn2016beyond,weng2016topological,zhu2016triple,chang2017nexus,singh2018topological,winkler2019topology,chapai2019fermions,shan2019new,wieder2016double}. Meanwhile, the classification can also be according to the dispersion around the band crossing. This is evidently an important characteristic, because it directly affects the properties of the low-energy excitations. The dispersion may have different leading order terms, e.g., with linear, quadratic, or cubic dispersions. Recent works have revealed the existence of stable quadratic and cubic nodal 
points as well as nodal lines~\cite{yang2014classification,fang2012multi,zhu2018quadratic,yu2019quadratic,wu2019higher,li2019type}. Here, in this article, we focus on the classification based on another aspect of the dispersion, which is specific for linear band crossings, namely, the signs of slopes of the crossing bands.

\begin{figure}[htbp]
	\includegraphics[width=8.5cm]{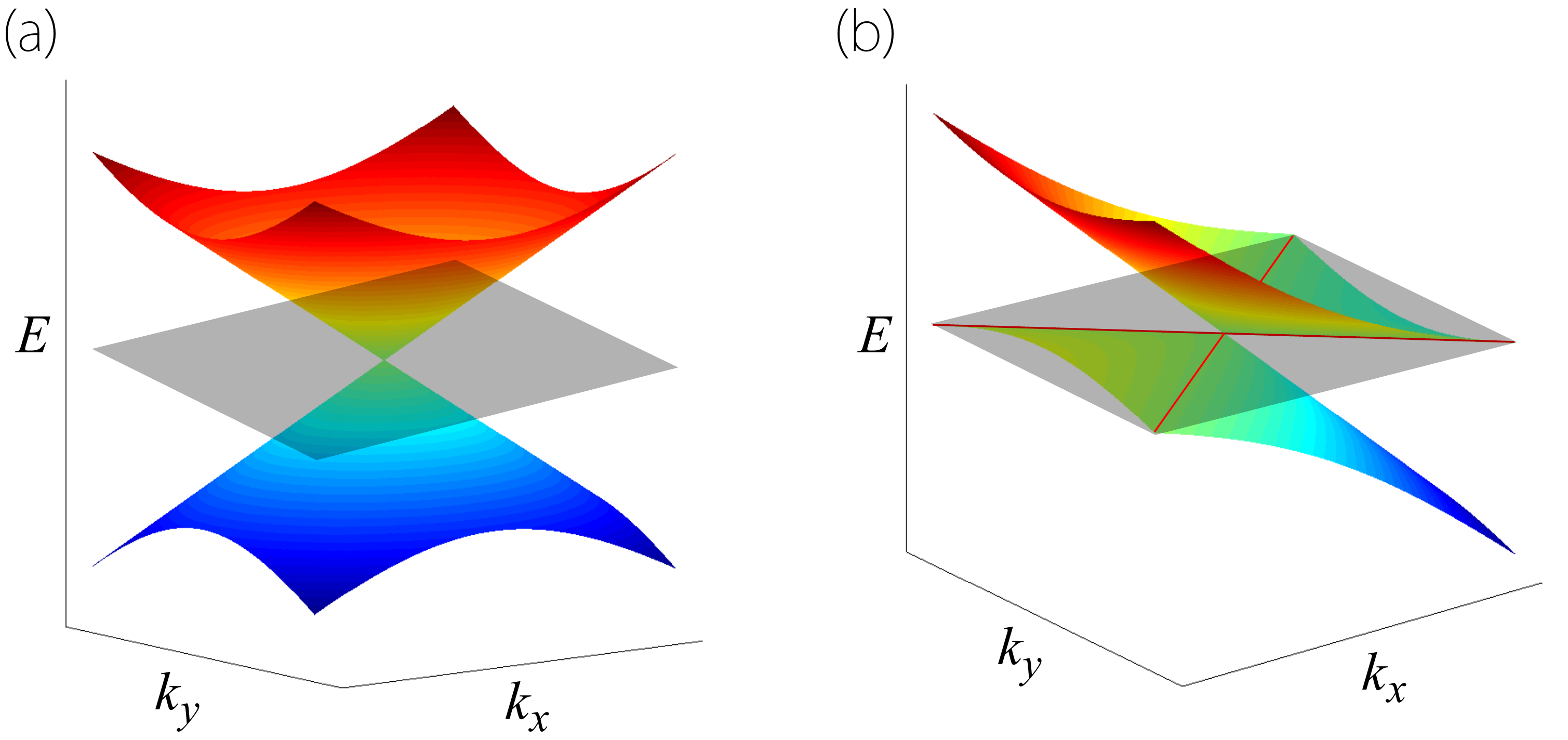}
	\caption{The schematic of the type-I (a) and type-II (b) Weyl points.}
	\label{fig1}
\end{figure}

This classification scheme was initially proposed for nodal points in 2015~\cite{soluyanov2015type,xu2015structured}. The dispersion around a conventional (type-I) Weyl point shows an up-right conical shape in the energy-momentum space [see Fig.~\ref{fig1}(a)]. In comparison, for the proposed type-II Weyl point, the cone is tipped over (i.e., the two crossing bands have the same sign for their slopes along certain directions in the $k$-space), as shown in Fig.~\ref{fig1}(b). The difference is crucial, because it leads to distinct Fermi surface topologies, which in turn determines many physical properties. In 2017, Li \emph{et al.}~\cite{li2017type} extends the scheme to nodal lines, proposing the concepts of type-II and hybrid nodal lines. These works have attracted great attention. Subsequent studies have reveal many interesting properties of these new TM phases, and have identified several realistic materials candidates.

In this article, we will briefly review the concepts, the physical properties, and the material realizations of the type-II nodal-point and nodal-line TMs. The very relevant concept of hybrid nodal-line TM and the speculation on the type-II nodal-surface TM will also be discussed. The field of TMs is still under rapid development. We hope this short review can serve its purpose to provide the readers with the background and to stimulate their interest to dive into this fascinating research field.

\section{Type-II nodal point}

We start by discussing the nodal-point TMs. The proposal of the type-II nodal point was first made for Weyl points~\cite{soluyanov2015type,xu2015structured}, and later extended to Dirac points~\cite{huang2016type,chang2017type}. In this section, we first review the concept. Then we discuss several interesting physical properties of the type-II nodal-point TMs. Finally, we review the material candidates that have been identified. It should be noted that the distinction between type-I and type-II is on the dispersion. They actually share the same kind of topology/symmetry protection. Here, we will not cover the fundamentals pertinent to the topology/symmetry aspects. The readers who are interested in these are suggested to consult the previous review articles in Refs.~\cite{armitage2018weyl,burkov2016topological,chiu2016classification,yang2016dirac,bansil2016colloquium}.

\subsection{Concept}

The concept of type-II dispersion was first introduced in the context of Weyl points. Here, we will follow the discussion in Ref.~\cite{soluyanov2015type} by Soluyanov \emph{et al.} The most general model describing an isolated Weyl point can be written as
\begin{equation}\label{WeylH}
H(\boldsymbol{k})=\sum_{i=x, y, z \atop j=0, x, y, z} k_{i} v_{i j} \sigma_{j},
\end{equation}
where the energy and the wave vector $\boldsymbol{k}$ are measured from the Weyl point, $k_i$ are the components of $\boldsymbol{k}$, $v_{ij}$ are the Fermi velocities, $\sigma_{0}$ is the $2 \times 2$ unit matrix, and $\sigma_{j}\ (j=x, y, z)$ are the three Pauli matrices. The energy dispersion around the point can be expressed as
\begin{equation}\label{WeylHeig}
\begin{split}
E_{\pm}(\boldsymbol{k})&=\sum_{i=x, y, z} k_{i} v_{i 0} \pm \sqrt{\sum_{j=x, y, z}\left(\sum_{i=x, y, z} k_{i} v_{i j}\right)^{2}}\\
&=T(\boldsymbol{k}) \pm U(\boldsymbol{k}).
\end{split}
\end{equation}
Here, $\pm$ corresponds to the two crossing bands. The two terms $T(\boldsymbol{k})$ and $U(\boldsymbol{k})$ both are linear in $k$. One notes that without the $T$ term, the spectrum has a chiral symmetry, i.e., it is symmetric with respect to $E=0$, as shown in Fig.~\ref{fig1}(a) where we have an up-right Weyl cone. A finite $T(\bm k)$ term will tilt the spectrum. When the tilt is so strong such that $T$ dominates over $U$ along a certain direction in $k$-space, the Weyl cone will be tipped over, and we will have a dramatically different situation, as shown in Fig.~\ref{fig1}(b). This is defined as the type-II Weyl point.

One observes that the transition from the conventional (type-I) Weyl point in Fig.~\ref{fig1}(a) to the type-II Weyl point in Fig.~\ref{fig1}(b) represents a Lifshitz transition: the Fermi surface for the former case is a point (the Weyl point), whereas for the latter case it consists of finite electron and hole pockets touching at the Weyl point. Since many physical properties are connected with the Fermi surface topology, one naturally expects that the type-II Weyl TMs would exhibit distinct physics from their type-I counterparts. This is the rationale behind the definition of type-II dispersion.

A more intuitive criterion for distinguishing type-I and type-II points is as follows. For a type-I point, along all directions in $k$-space, the two crossing bands always have slopes with opposite signs. In contrast, for a type-II point, there must exist a direction $\hat{\bm{k}}$, along which the two bands have slopes with the same sign. For example, in Fig.~\ref{fig1}(b), the two bands both have negative slopes along the $k_x$ direction, and hence the Weyl cone is tipped over along this direction. It should be noted that along other directions (such as the $k_y$ direction in Fig.~\ref{fig1}(b)), the two bands may still have slopes with opposite signs. This indicates that the type-II points intrinsically have a very strong anisotropy.

The concept can be naturally extended to Dirac points with preserved $\mathcal{PT}$ symmetry~\cite{huang2016type,chang2017type}, where $\mathcal{P}$ and $\mathcal{T}$ refer to the inversion and time reversal symmetries, respectively. Most discovered Dirac TM materials indeed have this symmetry (some exceptions have also been found~\cite{chen2017ternary,hu2019three}). With $\mathcal{PT}$ symmetry, each band is doubly degenerate, so the dispersion around a Dirac point is just like that for a Weyl point, and the above discussion directly applies.

\subsection{Physical property}

As mentioned above, the type-II nodal point is associated with a distinct Fermi surface topology from the type-I case. For the type-I case, we have a single Fermi point (when the Fermi energy is exactly at the nodal point) or a single carrier pocket, whereas for the type-II case, we have coexisting electron and hole pockets close to each other. This distinction leads to marked differences in their physical properties.

\begin{figure}[htbp]
	\centering
	\includegraphics[width=8.5cm]{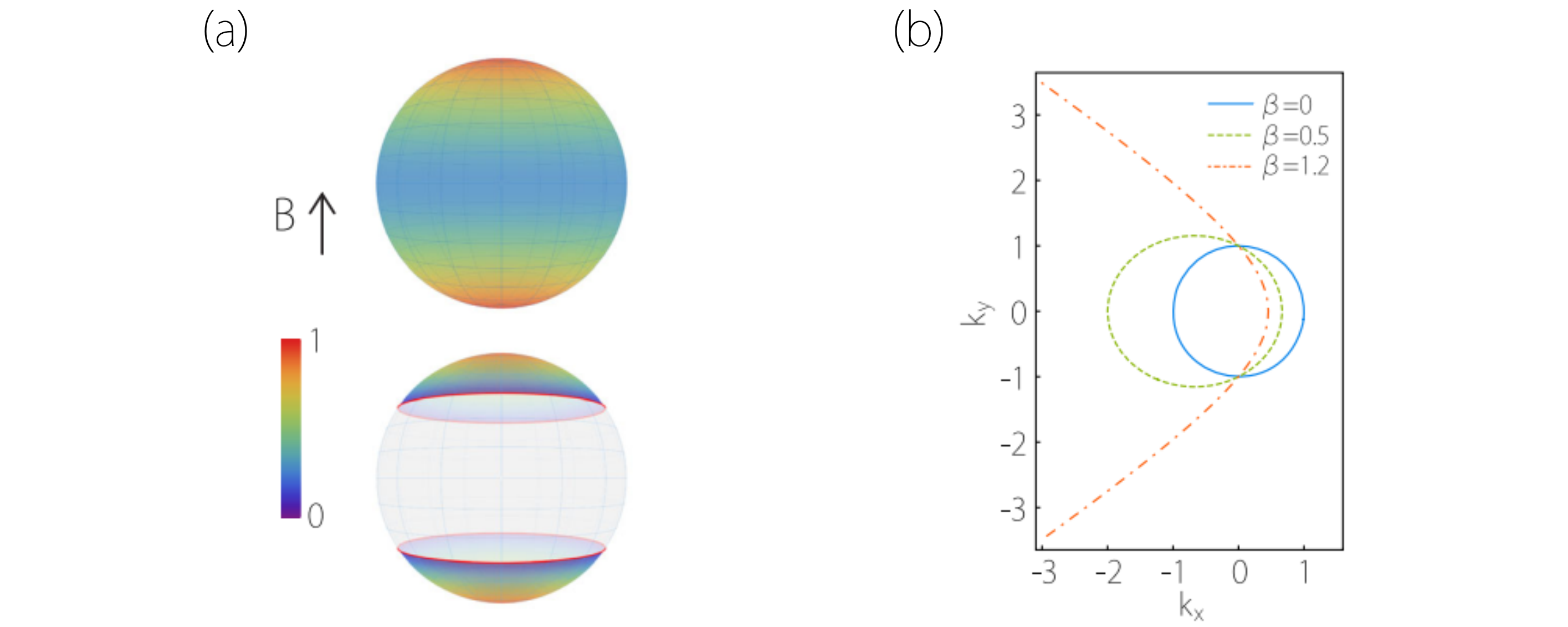}
	\caption{ (a) Landau level squeezing factor plotted versus tilt direction on a unit sphere. (Upper): type-I Weyl node. (Lower): Type-II node, in which the two red loops mark the critical angle where the Landau levels collapse. (b) Semiclassical orbit transforms from closed orbit to open trajectory for type-II Weyl node. Reproduced with permission from Ref.~\cite{yu2016predicted}.}
	\label{fig2}
\end{figure}

The Fermi surface topology will directly affect the magnetic response. Under an applied magnetic field, the electrons in solids will form cyclotron orbits. In $k$-space, these orbits are located on equi-energy surfaces and are within planes perpendicular to the field direction. Thus, the magnetic response is sensitive to the Fermi surface and offers an well established tool for probing the Fermi surface property. In Ref.~\cite{yu2016predicted}, Yu \emph{et al.} compared the Landau spectra for type-I and type-II points (see Fig.~\ref{fig2}). They found that the tilt [the $T$ term in Eq.~(\ref{WeylHeig})] tends to squeeze the Landau levels in energy. For a type-II point, there exists a critical angle between the $B$ field and the tilt direction, at which the squeezing becomes so strong such that the Landau levels collapses onto each other, leading to a huge peak in the density of states. Remarkably, this effect happens regardless of the magnetic field strength.
Similar Landau level collapse effect was also predicted for graphene~\cite{lukose2007novel}, but in that case, a huge in-plane electric field is also required, which proves to be challenging to achieve in experiment. In comparison, the effect is much more accessible for type-II TMs. It can manifest in the scanning tunneling spectroscopy (STS) and the magneto-transport measurements. Indeed, promising evidences have been reported in experiment on the type-II Weyl TM material WTe$_2$~\cite{li2017evidence}.

\begin{figure}[htbp]
	\centering
	\includegraphics[width=8.5cm]{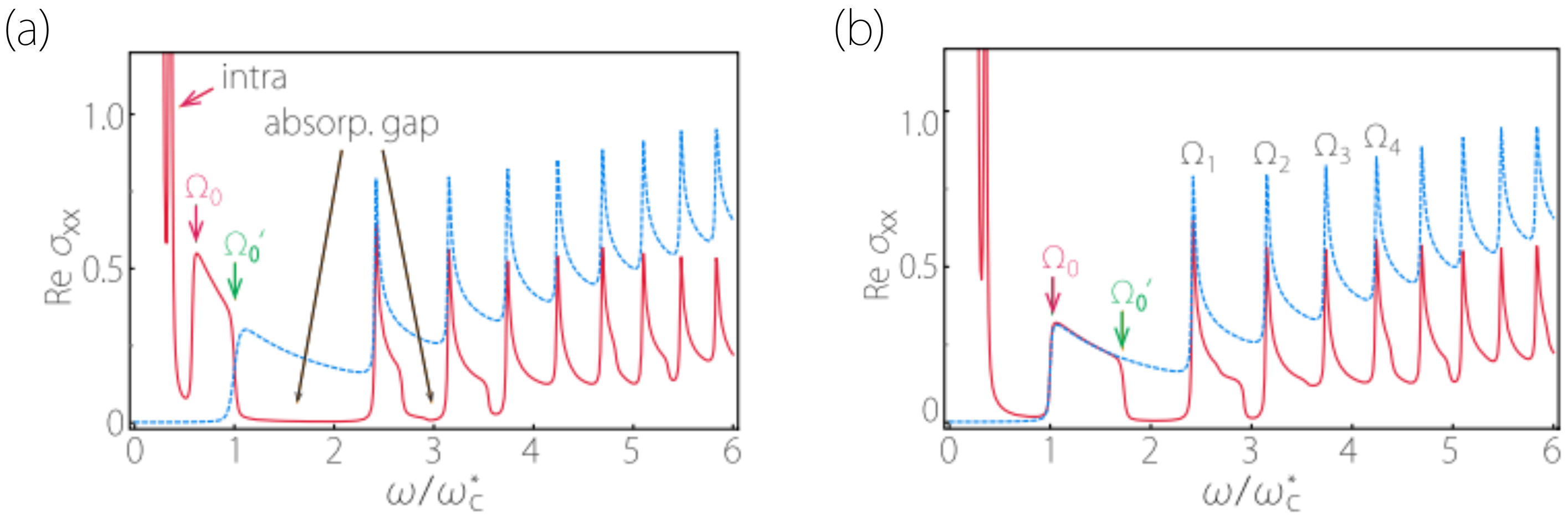}
	\caption{ (a) Optical conductivity for the type-I (blue dashed curve) and type-II (red solid curve) Weyl node. (b) is the same as (a) but with a reversed field direction. Reproduced with permission from Ref.~\cite{yu2016predicted}.}
	\label{fig3}
\end{figure}

Before the Landau level collapse (i.e., below the critical angle), there also exist distinct features in the magneto-optical response. As discussed in Refs.~\cite{yu2016predicted,tchoumakov2016magnetic,udagawa2016field}, these features include the invariable presence of intraband absorption peaks at low frequencies, the absence of absorption tails (as evidenced by the appearance of absorption gaps in Fig.~\ref{fig3}), and the special anisotropic field dependence. Especially, when reversing the $B$ field direction,
all the peaks are unchanged except for the first interband peak (which involves the chiral Landau band). This difference is illustrated in Fig.~\ref{fig3}, and is unique for the type-II points. In
Ref.~\cite{tchoumakov2016magnetic}, Tchoumakov \emph{et al.} further emphasized the existence of new types of optical transitions beyond the unusual dipolar ones. These features should be detectable by using the magneto-optical spectroscopy.

\begin{figure}[htbp]
	\centering
	\includegraphics[width=8.5cm]{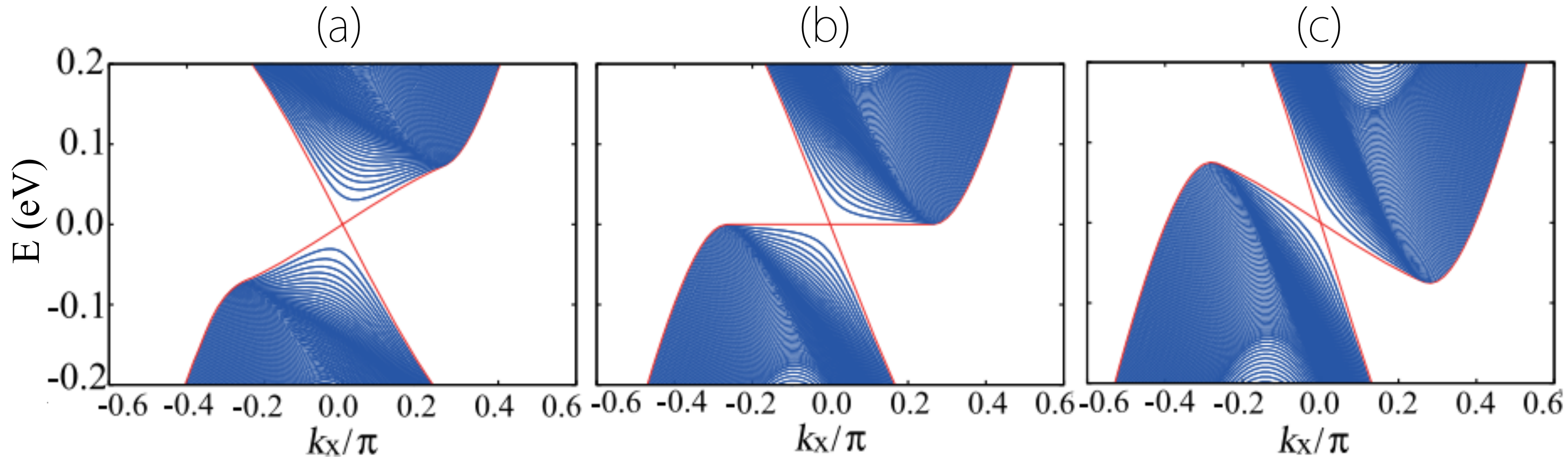}
	\caption{Landau spectra are obtained for increasing tilt term and (c) is the type-II. Reproduced with permission from Ref.~\cite{udagawa2016field}.}
	\label{fig4}
\end{figure}

A celebrated effect discussed in the context of Weyl semimetals is the chiral anomaly~\cite{nielsen1983adler,son2013chiral}. Under a magnetic field, there appears a special chiral Landau band around an original Weyl point, which is gapless and has its sign of slope determined by the chirality of the Weyl point. In a non-interacting lattice model, the Weyl points come in pairs of opposite chirality. This is known as the Nielson-Ninomiya no-go theorem~\cite{nielsen1981absence,nielsen1981absenceb} (one way to circumvent this theorem was recently proposed in Ref.~\cite{yu2019circumventing}). As a result, with $B$ field, the chiral Landau bands must also appear in pairs with slopes of opposite signs. If we further apply an electric field \emph{parallel} to the magnetic field, the electrons can be pumped through these chiral Landau bands from Weyl points of one chirality to those with the opposite chirality. This non-conservation of chirality corresponds to the chiral anomaly originally discussed in high energy physics. In condensed matter, this effect leads to a positive contribution to the longitudinal magneto-conductivity~\cite{nielsen1983adler,son2013chiral}. (However, positive longitudinal magneto-conductivity cannot be regarded as a smoking-gun evidence for the Weyl TM phase, because there exist other possible mechanisms even in the absence of nontrivial band topology~\cite{gao2017intrinsic,goswami2015axial}.). In Ref.~\cite{soluyanov2015type}, Soluyanov \emph{et al.} pointed out that the chiral anomaly has a strong dependence on the magnetic field direction. It occurs
only when angle between the $B$ field and the tilt is less than the critical angle mentioned above (or equivalently, when
$|T(\boldsymbol {k})|>|U(\boldsymbol {k})|$ for $\bm k$ along the field direction). Later, in Ref.~\cite{udagawa2016field}, by calculations on a lattice model for a type-II Weyl metal, it was shown that an inversion-asymmetric over-tilting can create an
imbalance in the number of chiral Landau bands with positive and negative slopes. This is clearly observed  in Fig.~\ref{fig4}(c), where the two chiral Landau bands both have negative slopes. So far, it is still not clear whether such chiral bands can generate any measurable signals in transport experiment.

\begin{figure}[htbp]
	\centering
	\includegraphics[width=8.5cm]{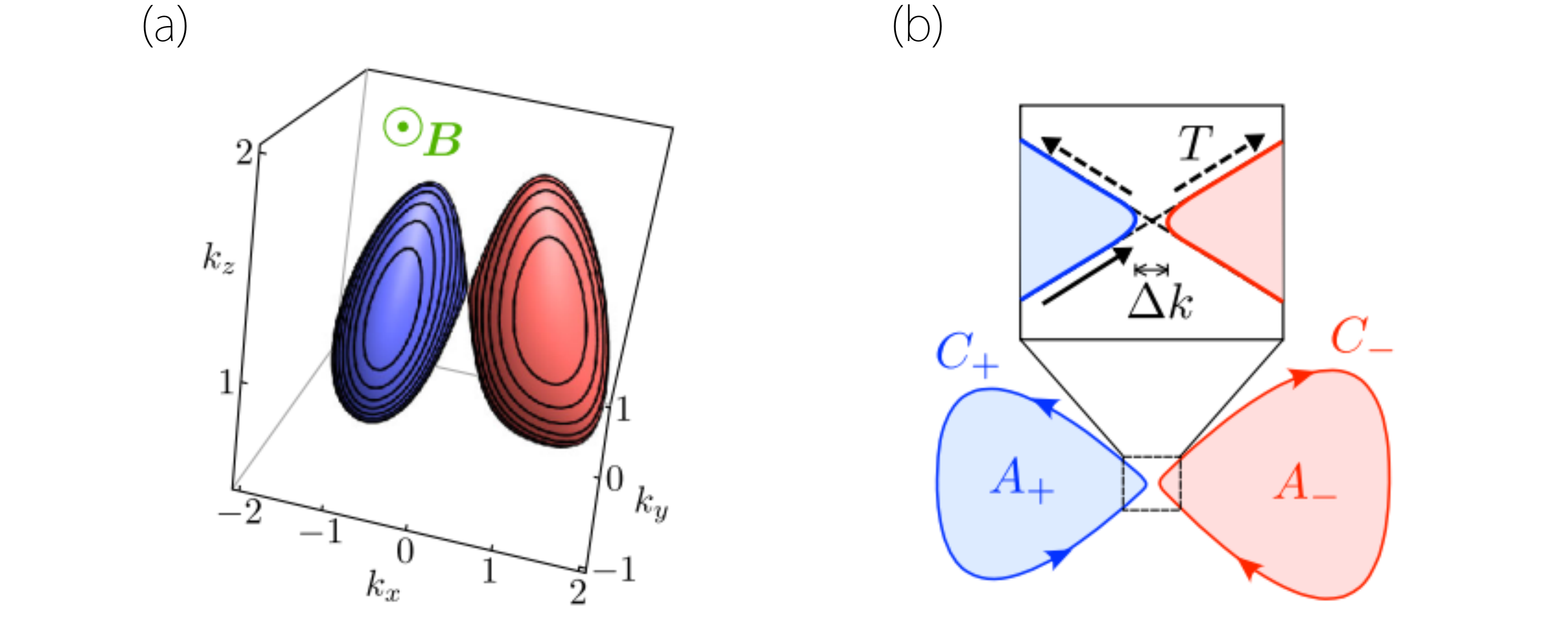}
	\caption{(a) Fermi surface of a type-II Weyl semimetal, showing the electron and hole pockets 
		touching at the Weyl point. Equienergy contours in planes perpendicular to the magnetic field $B$ are indicated. (b) Intersection of the Fermi 
		surface with a plane perpendicular to $B$ that passes near the Weyl point. Electron and hole pockets are bounded by a contour $C_{\pm}$ enclosing an area $A_{\pm}$. Reproduced with permission from Ref.~\cite{oBrien2016magnetic}.}
	\label{fig5}
\end{figure}

The energy spectrum under a magnetic field can usually be well captured by the semiclassical picture, involving the semiclassical cyclontron orbits quantized by the Bohr-Sommerfeld quantization condition~\cite{sundaram1999wave,xiao2010berry,cai2013magnetic}. This is also explicitly demonstrated for the Weyl model case~\cite{yu2016predicted}. One quantum effect going beyond this theory is the magnetic breakdown, referring to the field-induced tunneling between the semiclassical orbits. As we have mentioned above, a type-II nodal point is a contact point between an electron pocket and a hole pocket. Hence, one can naturally expect that there could exists strong tunneling between the electron and hole orbits. This effect has been studied in detail in Refs.~\cite{oBrien2016magnetic,koshino2016cyclotron}. As shown in Fig.~\ref{fig5}, the tunneling leads to hybridized 8-shaped orbits. O'Brien \emph{et al.}~\cite{oBrien2016magnetic} made an insightful comparison of this effect with the Klein tunneling at a $p$-$n$ junction in real space, and showed that as the energy approaches the Weyl point, the tunneling probability approaches unity even with vanishing $B$ field strength. This magnetic breakdown can manifest as an additional low frequency component in the magnetic oscillations. Koshino~\cite{koshino2016cyclotron} showed that the 8-shaped orbits also lead to a factor of two difference in the resonance peaks for the optical conductivities along two orthogonal directions normal to the $B$ field.

\begin{figure}[htbp]
	\centering
	\includegraphics[width=8.5cm]{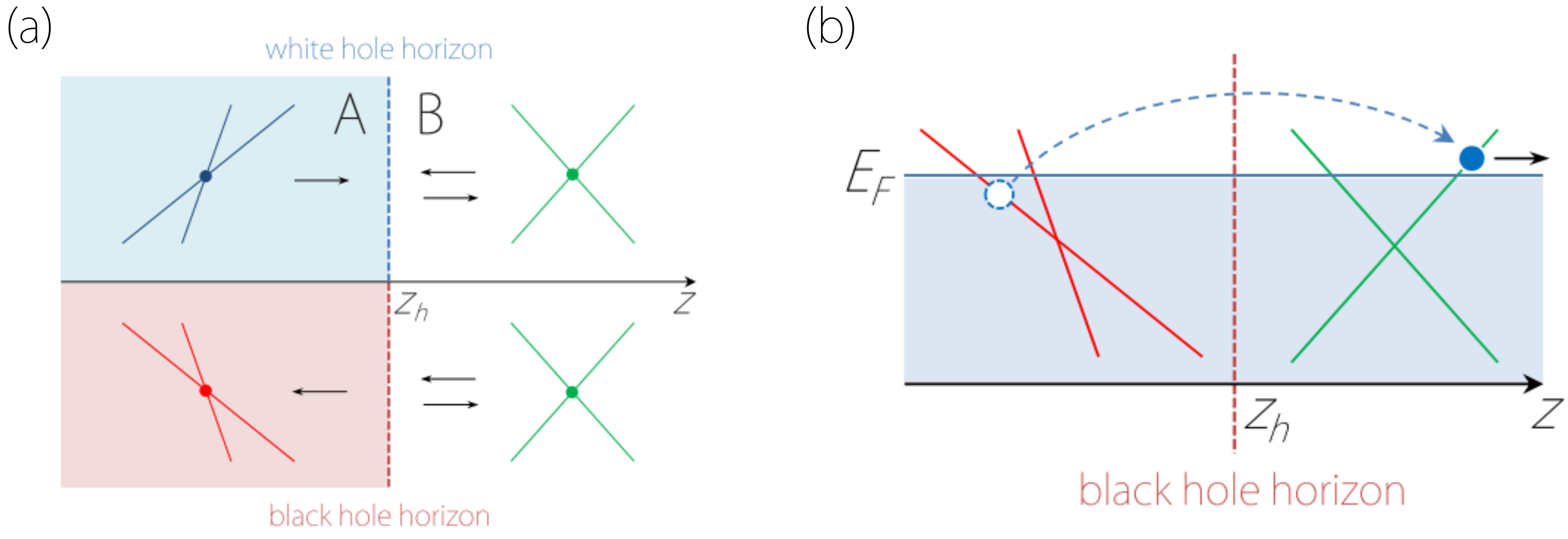}
	\caption{(a) Schematic figure showing (up) a white-hole horizon and (bottom) a blackhole horizon at $z=z_h$, corresponding to the Schwarzschild radius of metric. (b) Analogue of Hawking radiation. Illustration of the analogue of Hawking radiation in a topological semimetal. Reproduced with permission from Ref.~\cite{guan2017artificial}.}
	\label{fig6}
\end{figure}

In general relativity and astrophysics, it is known that the light-cone is tilted by the effects of gravity. And inside the event horizon of a black hole, the light-cone is tipped over, just like the type-II case here. This had inspired the early proposals to realize artificial black holes in the superfluid Helium~\cite{unruh1981experimental,volovik2003universe}, via controlling vortices and background superfluid flow which are challenging in experiment. Now, the type-II TMs offer a new possibility. In Ref.~\cite{guan2017artificial}, Guan \emph{et al.} proposed to achieve several astrophysical analogues in strained Weyl/Dirac TMs, including the black/white hole event horizons, gravitational lens, and Hawking radiation. It was shown that an event horizon can be created as a domain wall between type-I and type-II regions [see Fig.~\ref{fig6}(a)]. A simple formula for estimating the Hawking radiation temperature was also presented.

Finally, for conventional type-I Weyl semimetals with broken $\mathcal{T}$ symmetry, it has been shown that there exists a finite intrinsic anomalous Hall effect with magnitude proportional to the separation between the Weyl points~\cite{yang2011quantum}. Zyuzin and Tiwari~\cite{zyuzin2016intrinsic} found that for the type-II case, the anomalous Hall conductivity has a sensitive dependence on the tilt. Moreover, the conductivity remains finite even when the separation between the Weyl points vanishes, if the tilts of the Weyl points are not identical.

\subsection{Material realization}

\begin{figure}[htbp]
	\centering
	\includegraphics[width=8.5cm]{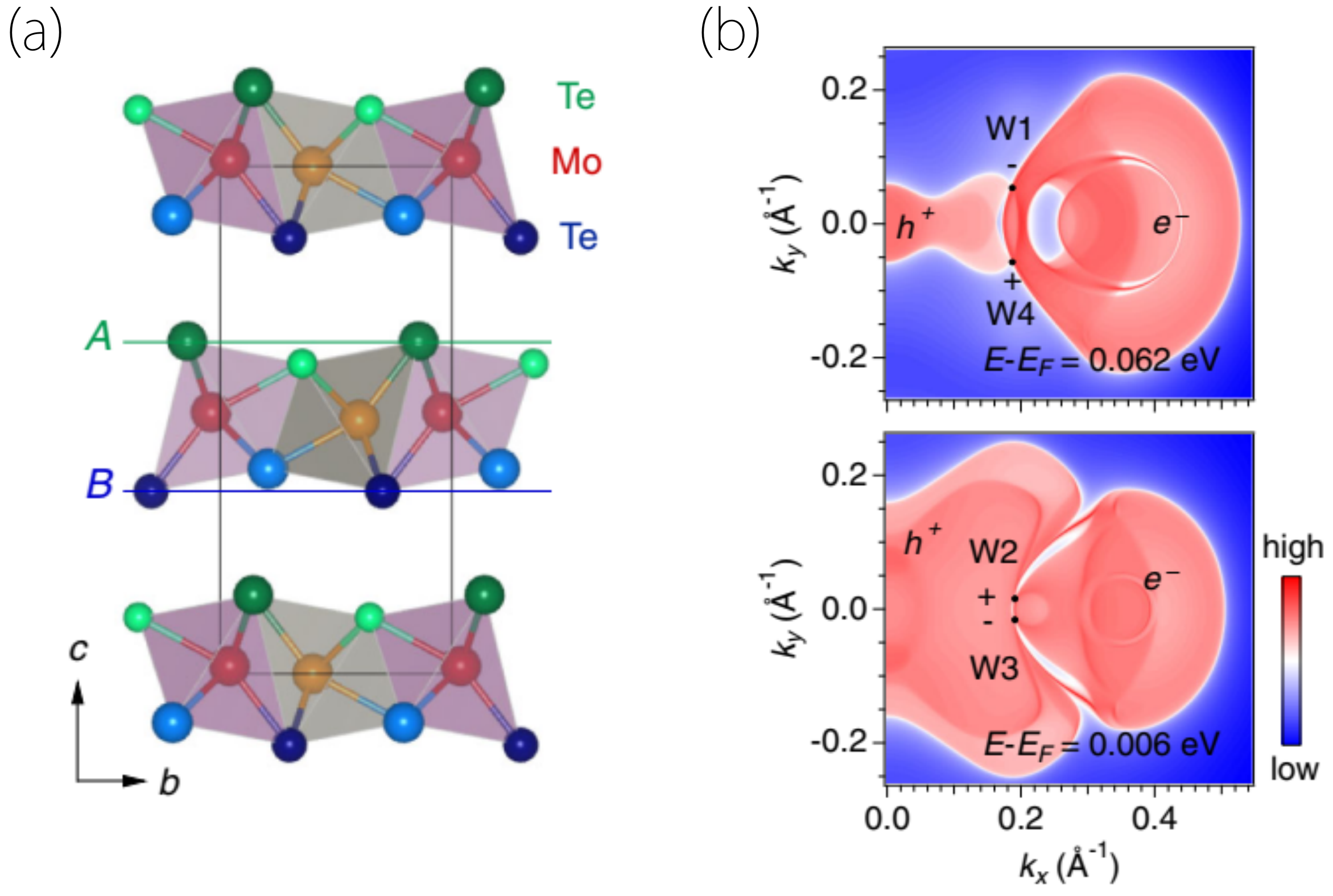}
	\caption{(a) Type-II Weyl TM in MoTe$_2$. (a) Crystal structure of the 1$\mathrm{T}^{\prime}$ phase of MoTe$_2$. (b) Momentum-resolved density of states of the (001) surface at two constant energies, showing the Fermi arcs connecting the projections of the type-II Weyl points. Reproduced with permission from Ref.~\cite{tamai2016fermi}}.
	\label{fig7}
\end{figure}

The first example of the type-II Weyl TM was found by Soluyanov \emph{et al.} in the layered material WTe$_2$~\cite{soluyanov2015type}.  Subsequently, a number of type-II Weyl TMs have been identified, including MoTe$_2$~\cite{jiang2017signature,deng2016experimental,huang2016spectroscopic,tamai2016fermi}, Mo$_{x}$W$_{1-x}$Te$_{2}$~\cite{belopolski2016discovery,belopolski2016fermi,zheng2016atomic}, TaIrTe$_4$~\cite{koepernik2016tairte,haubold2017experimental}, LaAlGe~\cite{xu2017discovery}, $X$P$_{2}$($X=$Mo, W)~\cite{autes2016robust,kumar2017extremely}, Ta$_{3}$S$_{2}$~\cite{chang2016strongly} and YbMnBi$_2$~\cite{borisenko2019time,zhu2019scanning}. Figure.~\ref{fig7} shows the example of MoTe$_2$. Although Weyl cones are completely tipped over along a certain direction. Type-II dispersion does not affect the topological charge of the Weyl point. Thus, a type-II Weyl TM also possesses Fermi arc surface
states, as shown in Fig.~\ref{fig7}(b) for MoTe$_2$~\cite{tamai2016fermi}.

\begin{figure}[htbp]
	\centering
	\includegraphics[width=8.5cm]{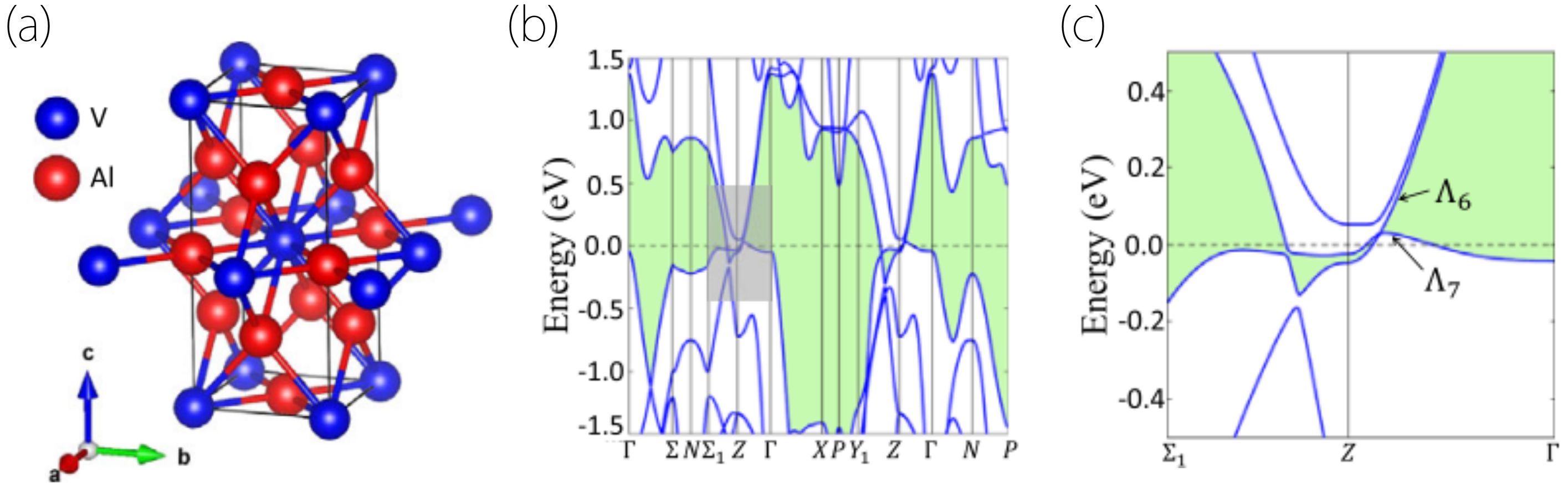}
	\caption{(a) The crystal structure of VAl$_3$. (b) The calculated bulk band structure of VAl$_3$ in the presence of SOC. The green shaded region shows the energy gap between the lowest conduction and valence bands. (c) An enlarged view of the area highlighted by the gray box in (b). Reproduced with permission from Ref.~\cite{chang2017type}.}
	\label{fig8}
\end{figure}

Many TMs with type-II Dirac points have also been discovered to date, such as PtTe$_{2}$~\cite{yan2017lorentz}, PtSe$_{2}$~\cite{huang2016type,zhang2017experimental}, PdTe$_{2}$~\cite{noh2017experimental,fei2017nontrivial}, $MA_{3}$ ($M=$V, Nb, Ta; $A=$Al, Ga, In)~\cite{chang2017type}, YPd$_2$Sn class~\cite{guo2017type},
LaAlO$_3$/LaNiO$_3$/LaAlO$_3$ quantum well~\cite{tao2018two}, high-temperature cuprate superconductor
$\mathrm{La}_{2-x} \mathrm{Sr}_{x} \mathrm{CuO}_{4}$ and $\mathrm{Eu}_{2-x} \mathrm{Sr}_{x}
\mathrm{NiO}_{4}$~\cite{horio2018two}, Heusler compounds $X$InPd$_2$ ($X=$Ti, Zr, Hf)~\cite{mondal2019type}, and NiTe$_2$~\cite{ghosh2019observation}. The crystal structure and the band structure of VAl$_3$ are shown in Fig.~\ref{fig8}. In Fig.~\ref{fig8}(c), one can clearly observe that the type-II crossing point on the $Z$-$\Gamma$ path. Here, each band has a double degeneracy due to the $\mathcal{PT}$ symmetry, so that the point is a type-II Dirac point.

\begin{figure}[htbp]
	\centering
	\includegraphics[width=8.5cm]{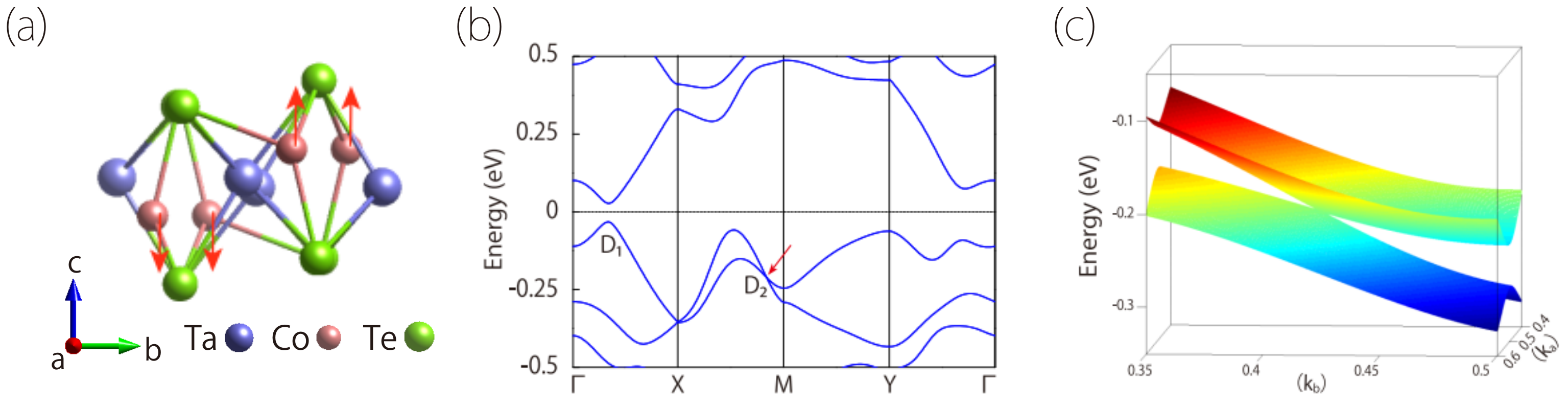}
	\caption{(a) The crystal structure of TaCoTe$_2$, the red arrows represent the direction of the magnetic moments. (b) Band structure of monolayer TaCoTe$_2$ with SOC included. The type-II Dirac point D$_2$ is indicated by the red arrow. (c) Two-dimensional band structure around the type-II Dirac point D$_2$. Reproduced with permission from Ref.~\cite{li2019two}.}
	\label{fig9}
\end{figure}

All the above mentioned type-II Dirac points are in non-magnetic materials. Recently, the first example of a magnetic type-II Dirac point was found in TaCoTe$_2$~\cite{li2019two}. This material is antiferromagnetic in its ground state, with the magnetic configuration shown in Fig.~\ref{fig9}(a). Again the system has a composite  $\mathcal{PT}$ symmetry, so each band is doubly degenerate.
In its band structure, one finds a type-II Dirac point slightly below the Fermi level, as indicated by the red arrow in Fig.~\ref{fig9}(b) Importantly, this Dirac point is robust against spin-orbit coupling.

\section{Type-II nodal line}

A nodal line can be viewed as consisting of many nodal points. Since the dispersion of a nodal point can be classified as
type-I or type-II, the following question naturally arises: can the classification be generalized to nodal lines? This question was first investigated by Li \emph{et al.} in 2017~\cite{li2017type}.

\subsection{Concept}

\begin{figure}[htbp]
	\centering
	\includegraphics[width=8.5cm]{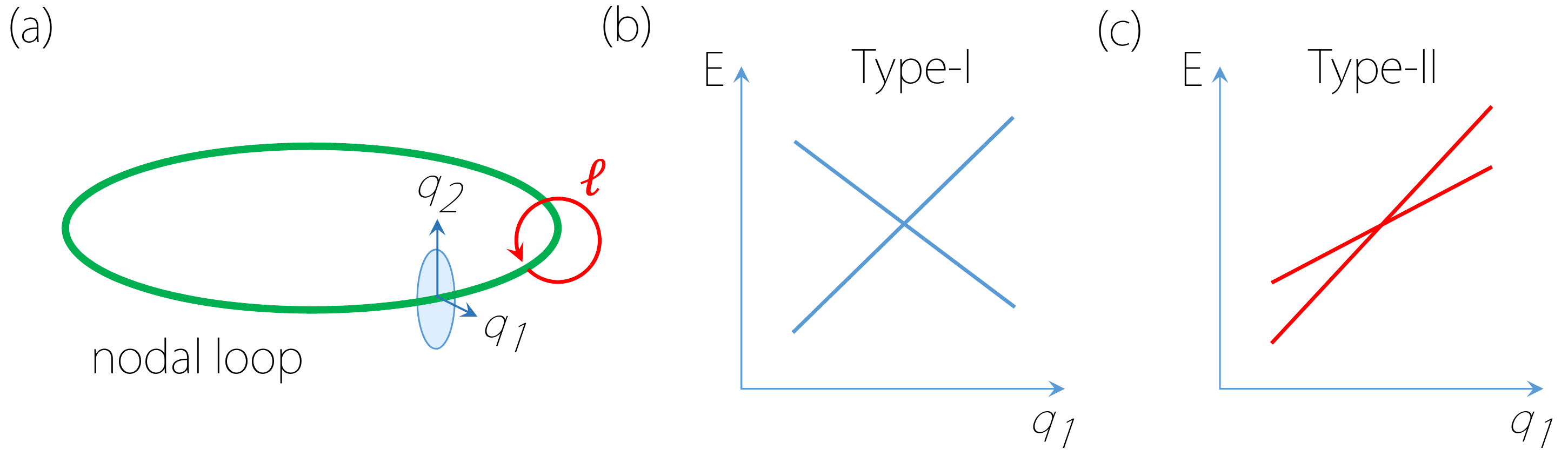}
	\caption{(a) Schematic figure of a nodal loop. $q_{1}$ and $q_{2}$ label the
		two transverse directions. (b) and (c) illustrate the type-I and type-II
		dispersions along the $q_{1}$ direction. Reproduced with permission from Ref.~\cite{li2017type}.}.
	\label{fig10}
\end{figure}

We first introduce the concept of a type-II nodal line. The discussion here follows that in Ref.~\cite{li2017type}. Consider a nodal line formed by the linear crossing between two bands in a 3D system. For any point $P$ on the line, the general low-energy effective model around $P$ can be expressed as
\begin{equation}\label{Pmodel}
\mathcal{H}(\bm q)=v_1 q_1 \sigma_x+v_2 q_2 \sigma_y+\boldsymbol  w\cdot\boldsymbol  q.
\end{equation}
Here, the wave-vector $\boldsymbol q$ is measured from $P$, $q_i$ ($i=1,2$) are the components of $\boldsymbol  q$ along the two orthogonal transverse directions [see Fig.~\ref{fig10}(a)], and $v_i$ are the Fermi velocities. The last term in (\ref{Pmodel}) represents the local tilt of the spectrum at $P$. The energy spectrum is given by
\begin{equation}\label{Epm}
E_{\pm}(\bm q)=\boldsymbol  w\cdot\boldsymbol  q\pm \sqrt{v_1^2 q_1^2+v_2^2 q_2^2}.
\end{equation}
It should be noted that in the $q_1$-$q_2$ plane, the tilt is most effective along the $\boldsymbol  w_\bot=(w_1,w_2,0)$  direction, where $\boldsymbol  w_\bot$ is the projection of $\boldsymbol  w$ onto the $q_1$-$q_2$ plane.
When $|\boldsymbol  w_\bot|$ is small, the slopes of two bands have opposite signs for all directions in the  $q_1$-$q_2$ plane and it is the conventional dispersion [Fig.~\ref{fig10}(b)], and the point $P$ can be defined as type-I.
However, when $|\boldsymbol  w_\bot|$ is large enough and satisfies the relationship $|\boldsymbol  w_\bot|^2>\sqrt{v_1^2 w_1^2+v_2^2 w_2^2}$, the spectrum becomes completely tipped along the $\boldsymbol  w_\bot$ direction [Fig.~\ref{fig10}(c)] and the slopes of the two crossing bands have the same sign. Such a dispersion (and the point $P$) can be referred to as type-II. Now, we can define a type-II (type-I) nodal line to be the one on which all the points are of type-II (type-I) dispersion. Generically, a type-I nodal line is formed by the crossing of an electron-like band and a hole-like band, whereas a type-II nodal line is formed by the crossing of two electron-like (or hole-like) bands. Hence, such a classification makes physical sense.

Ref.~\cite{li2017type} also presented the following minimal model for a type-II nodal line.
\begin{equation}\label{Hring}
H(\bm k)=\frac{1}{2m}k_\rho^2+\frac{1}{2\eta}(k_\rho^2-k_0^2)\sigma_x+v_zk_z\sigma_y,
\end{equation}
where $k_\rho=\sqrt{k_x^2+k_y^2}$, $m$, $\eta$, and $k_0$ are model parameters. This model describes a nodal loop with radius $k_0$ in the $k_z=0$ plane. Evidently, model (\ref{Pmodel}) can be regarded as its low-energy model via identifying $q_1$ to be along the $\hat{k}_\rho$ direction, $q_2$ to be along $\hat k_z$-direction, with the correspondences that $v_1=k_0/\eta$, and $\boldsymbol  w=k_0/m\hat k_\rho$. One finds that the model (\ref{Hring}) describes a type-II (type-I) nodal loop when $|\eta/m|>1$ ($<1$). For the type-II case (with $|\eta/m|>1$), the tilt term (first term in Eq.~(\ref{Hring})) dominates, making both bands electron-like or hole-like along the radial direction depending on the sign of $m$, and their intersection makes the type-II nodal loop. This model can be used as a start point for theoretical studies of the properties of type-II nodal lines.

\subsection{Physical property}

\begin{figure}[htbp]
	\centering
	\includegraphics[width=8.5cm]{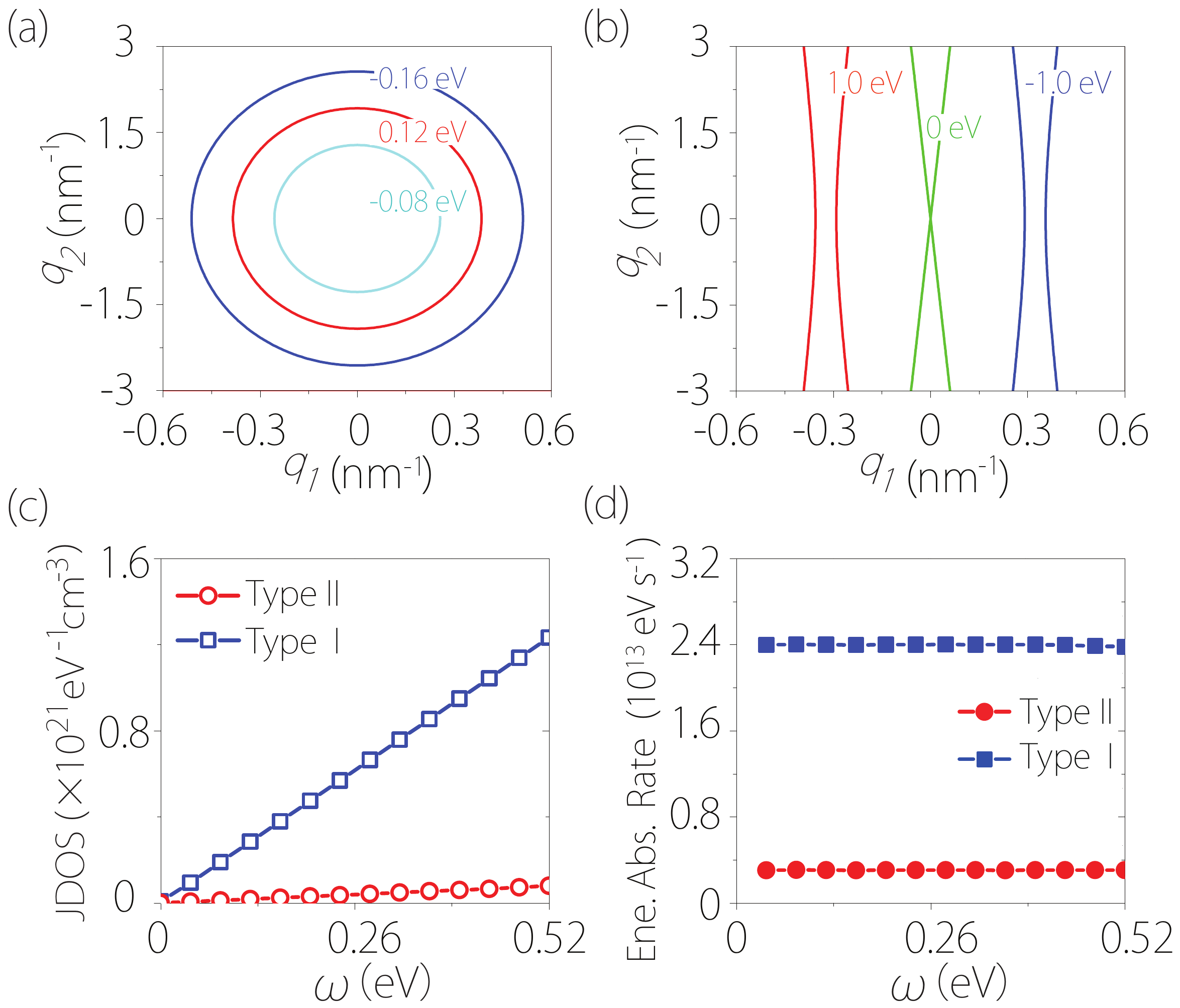}
	\caption{(a) Equienergy contours in the $q_{1}-q_{2}$ plane for (a) type-I and
		(b) type-II loops. Comparison of $(\mathrm{c})$ JDOS and (d) optical energy absorption rate for the two types of loops. Reproduced with permission from Ref.~\cite{li2017type}.}
		\label{fig11}
\end{figure}

Type-II and type-I nodal loops differ in their energy dispersions. Most importantly,
they have different kinds of constant energy surfaces. As shown in Fig.~\ref{fig11}(a,b) in the $q_1$-$q_2$ plane for model ($k_\rho$-$k_z$ plane for model), for type-I line, the equi-energy contours are closed ellipses, whereas for type-II line, the equi-energy contours become hyperbolas.
This qualitative difference will manifest in a variety of physical properties. For example, under a magnetic field, electrons orbit around constant energy surfaces, the different types of orbits will produce contrasting signals in magneto-oscillations~\cite{oBrien2016magnetic,li2016hybrid,khim2016magnetotransport} (such as de Haas-van Alphen oscillations); and the transition from elliptic to hyperbolic type orbits, e.g., by varying the magnetic field direction for a type-II nodal loop, would typically be accompanied by the Landau level collapse effect~\cite{yu2016predicted} as discussed in Sec.~II(B).

In Fig.~\ref{fig11}(a,b), one also observes that the positive and negative energy contours have a much less overlap for a type-II nodal line than for a type-I nodal line. This is a natural consequence of both bands being electron-like (or hole-like), and could lead to marked difference in their optical response. In Fig.~\ref{fig11}(c,d), the joint density of states (JDOS): $\mathcal{D}(\omega)=\frac{1}{V}\sum_{\boldsymbol  k}\delta(E_{c,\boldsymbol  k}-E_{v,\boldsymbol  k}-\omega)$, and the optical energy absorption rate: $\mathcal{R}(\omega)={2\pi}\omega\sum_{\boldsymbol  k}|M_{cv}|^2 \delta(E_{c,\boldsymbol  k}-E_{v,\boldsymbol  k}-\omega)$ are presented for the type-II and type-I nodal loops. Here $M_{cv}$ is the optical transition matrix element and the model in Eq.~(\ref{Hring}) is used for the calculation. One indeed observes that both quantities are much suppressed for the type-II nodal line.

For carrier transport in the plane of a nodal loop, the type-II loop should have a higher mobility than the type-I loop. This is because the low-energy states near a point on the type-II loop are propagating roughly in the same direction, while the opposite-propagating states are located at the other end of the loop (see Fig.~\ref{fig10}), thus momentum relaxation by scattering would be less effective as compared with the type-I case~\cite{guan2017artificial}.

In addition, Chen \emph{et al.}~\cite{chen2018floquet} found that under light-irridiation, a type-II nodal loop shows interesting transformations into several Floquet Weyl semimetal states, with features distinct from a type-I loop.

\subsection{Material Realization}

\begin{figure}[htbp]
	\centering
	\includegraphics[width=8.5cm]{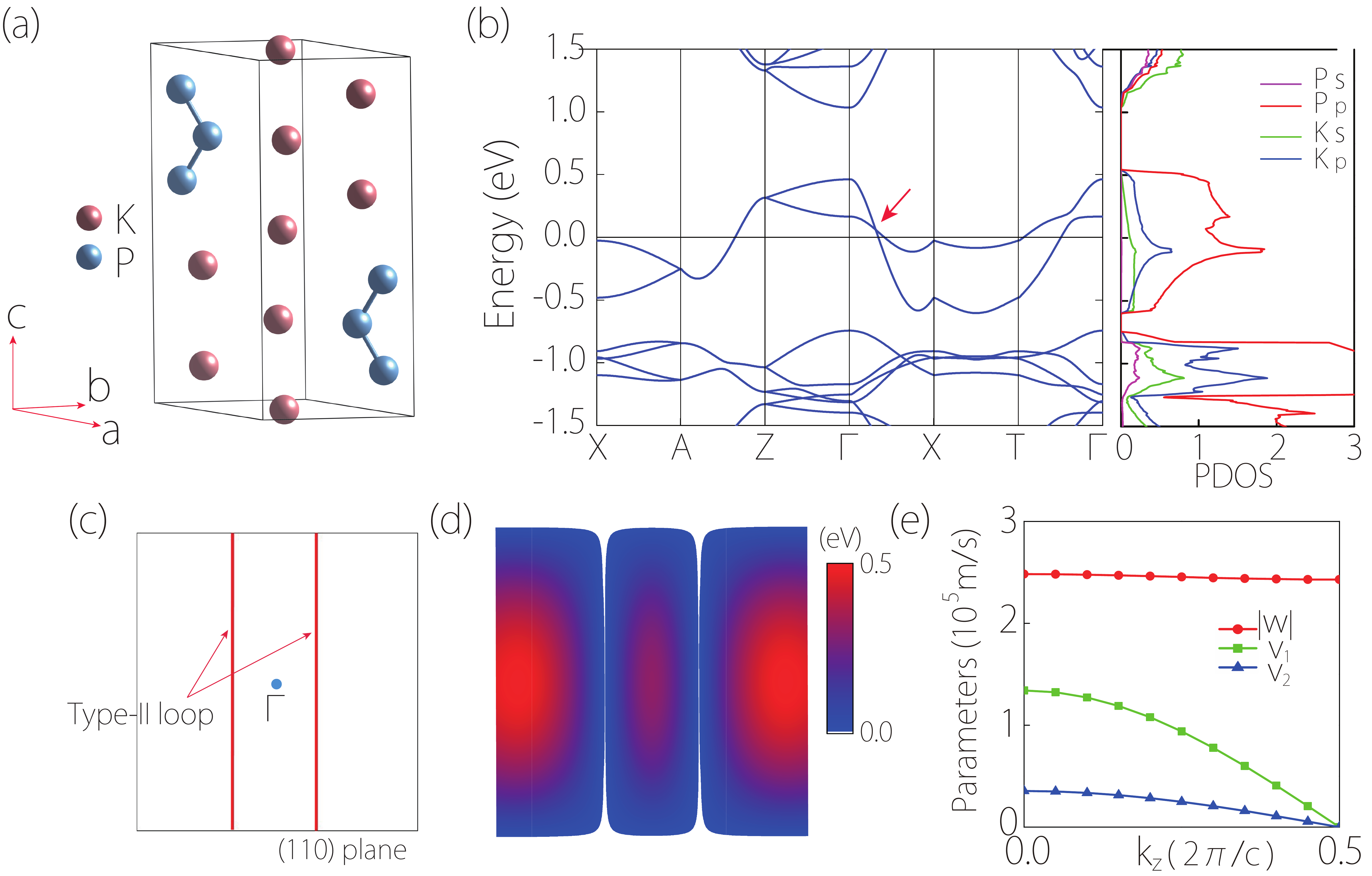}
	\caption{(a) The crystal structure of K$_4$P$_3$. (b) Electronic band structure of K$_4$P$_3$ and the projected density of states (PDOS). The red arrow indicates the crossing point on a type-II nodal loop. (c) Schematic figure showing the location of the type-II loops in the (110) plane, and (d) shows the corresponding result from DFT. The color-map shows the local gap between the two crossing bands. (e) Parameters of effective model (\ref{Pmodel}) obtained by fitting the DFT band structure. Reproduced with permission from Ref.~\cite{li2017type}.}
	\label{fig12}
\end{figure}

The first material example with a type-II nodal line was predicted in the crystalline compound K$_4$P$_3$~\cite{li2017type}. The material has the W$_3$CoB$_3$-type orthorhombic structure with space group No.~63 (\emph{Cmcm}) [see Fig.~\ref{fig12}(a)].
Figure ~\ref{fig12}(b) shows the band structure of K$_4$P$_3$, in which one can observe a linear band crossing point along the $\Gamma$-X paths very close to the Fermi level, as indicated in Fig.~\ref{fig12}(b). This linear crossing point is not isolated, but belongs to a nodal loop traversing the Brillouin zone (such a shape features a nontrivial $\mathbb{Z}^3$ index as discussed in Ref.~\cite{li2017type}), as illustrated in Fig.~\ref{fig12}(c,d). Furthermore, the dispersion in Fig. is of type-II. One can use the model (\ref{Pmodel}) to fit the DFT band structure. The fitted parameters are shown in Fig.~\ref{fig12}(e). One observes that $|w|>|v_1|$ for the whole loop, which confirms that the loop is type-II.

\begin{figure}[htbp]
	\centering
	\includegraphics[width=8.5cm]{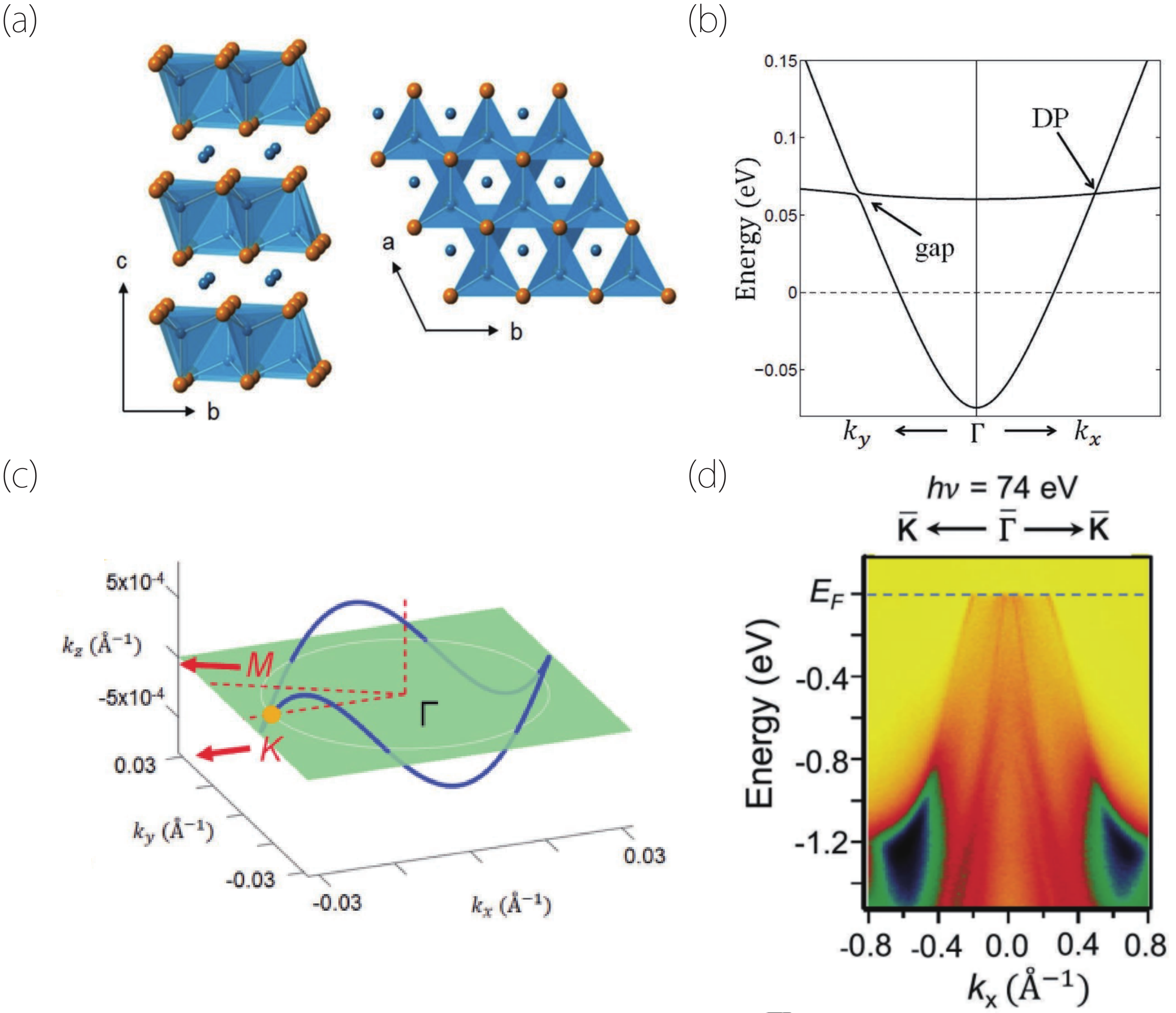}
	\caption{(a) The crystal structure of Mg$_3$Bi$_2$. (b) Type-II nodal-line band structure of Mg$_3$Bi$_2$ (without SOC). (c)
		The 3D plot of the nodal line of  Mg$_3$Bi$_2$ with a closed loop surrounding the $\Gamma$ point. (d) ARPES spectrum taken along the $\bar{\Gamma}$-$\bar{K}$ direction. The photon energy is 74 eV. Reproduced with permission from Ref.~\cite{chang2019realization}.}
	\label{fig13}
\end{figure}

Subsequently, type-II nodal lines have also been found in a few other materials, such as Mg$_3$Bi$_2$~\cite{zhang2017topological,chang2019realization} and strained Na$_3$N~\cite{kim2018type}. Figure~\ref{fig13} shows the band structure results of Mg$_3$Bi$_2$. The type-II nodal loop in Mg$_3$Bi$_2$ has been verified by the 
angle-resolved photoemission spectroscopy (ARPES) measurement~\cite{chang2019realization}.

\begin{figure}[htbp]
	\centering
	\includegraphics[width=8.5cm]{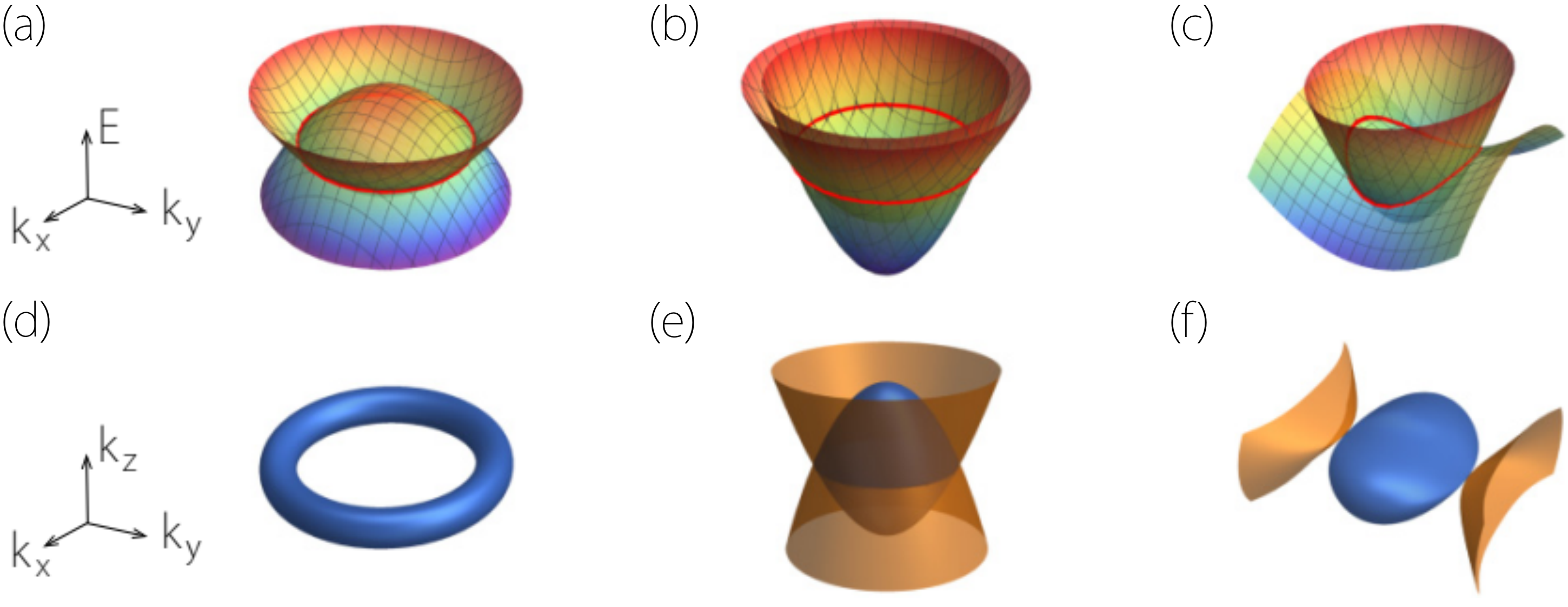}
	\caption{Illustration of (a) type-I, (b) type-II, and (c) hybrid nodal loops, with their typical Fermi surfaces shown in (d)-(f),
		respectively. Here, the loop is assumed to be in the $k_{x}$-$k_{y}$ plane, and blue (orange) color in (d)-(f) indicates the electron
		(hole) character of the Fermi surface. Reproduced with permission from Ref.~\cite{zhang2018hybrid}.}
	\label{fig14}
\end{figure}

\section{Hybrid nodal line}

Besides type-I and type-II lines, there exists a remaining possibility for a nodal line: the line could be composed of both type-I and type-II points. This is defined as a hybrid nodal line. On a hybrid nodal line, the tilt dominates only part of the line.

A minimal model for a hybrid nodal line is presented in Ref.~\cite{zhang2018hybrid}:
\[
H(\bm k)=\alpha_{x} k_{x}^{2}+\alpha_{y} k_{y}^{2}+\left(\beta_{x} k_{x}^{2}+\beta_{y} k_{y}^{2}-M\right) \sigma_{x}+v_{z} k_{z} \sigma_{y}.
\]
Here, $\alpha_{i}, \beta_{i}, M,$ and $v_{z}$ are model parameters. Assuming
$\beta_{x(y)}>0$, when $M>0$, the model describes a nodal loop lies in the
$k_{x}-k_{y}$ plane, and around each point on the loop, the expanded low-energy model takes the form in Eq.~(\ref{Pmodel}). One can verify that when $\left|\alpha_{i}\right| /\left|\beta_{i}\right|<1(>1)$ $(i=x, y)$, the loop is type I (type II); and when $\left(\left|\alpha_{x}\right|-\right.$ $\left.\left|\beta_{x}\right|\right)\left(\left|\alpha_{y}\right|-\left|\beta_{y}\right|\right)<0$, a hybrid loop is realized.

\begin{figure}[htbp]
	\centering
	\includegraphics[width=8.5cm]{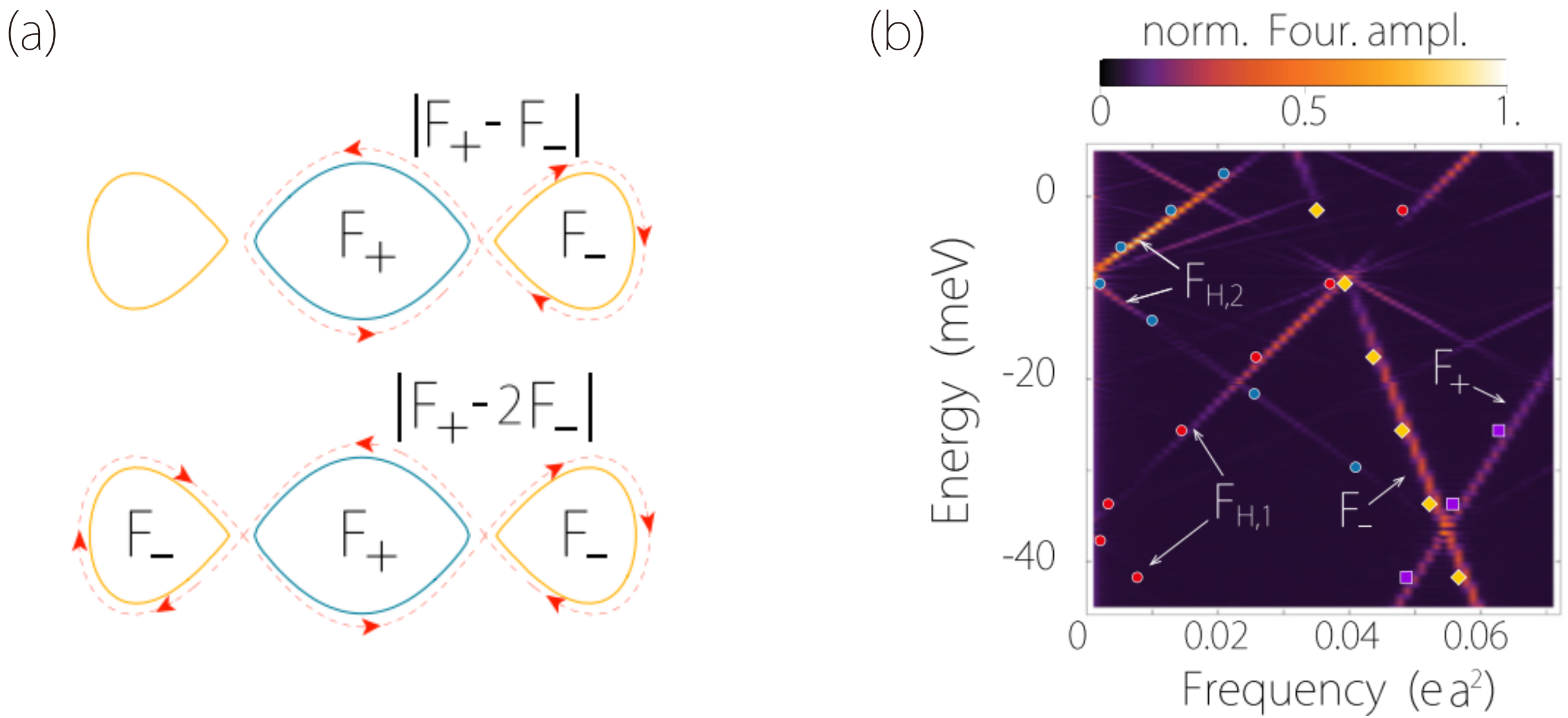}
	\caption{(a) Two types of hybridized orbits for $B$-field along the $x$-direction, with characteristic magneto-oscillation frequencies $\left|F_{+}-F_{-}\right|$ and $\left|F_{+}-2 F_{-}\right|$ respectively. (b) Fourier amplitudes of magnetic quantum oscillations for $B$-field along the $x$-direction. Reproduced with permission from Ref.~\cite{zhang2018hybrid}.}
	\label{fig15}
\end{figure}

The band structures as well as the Fermi surfaces for the type-I, type-II, and hybrid nodal lines are schematically illustrated in Fig.~\ref{fig14}. A hybrid loop would typically occur when one of the crossing bands has a saddle-type dispersion [see Fig.~\ref{fig14}(c)]. This leads to two important features. The first feature is that a hybrid loop generally has energy variation along the loop, i.e., it spans a range of energy. The second features is that when the Fermi energy is located in that energy range, the Fermi surface would consist of coexisting electron and hole pockets that are connected by isolated nodal points on the loop. These features can give rise to interesting magnetic responses, such as the zero-field magnetic breakdown (see Fig.~\ref{fig15}) and peculiar anisotropy in the cyclotron resonance~\cite{zhang2018hybrid}.

\begin{figure}[htbp]
	\centering
	\includegraphics[width=8.5cm]{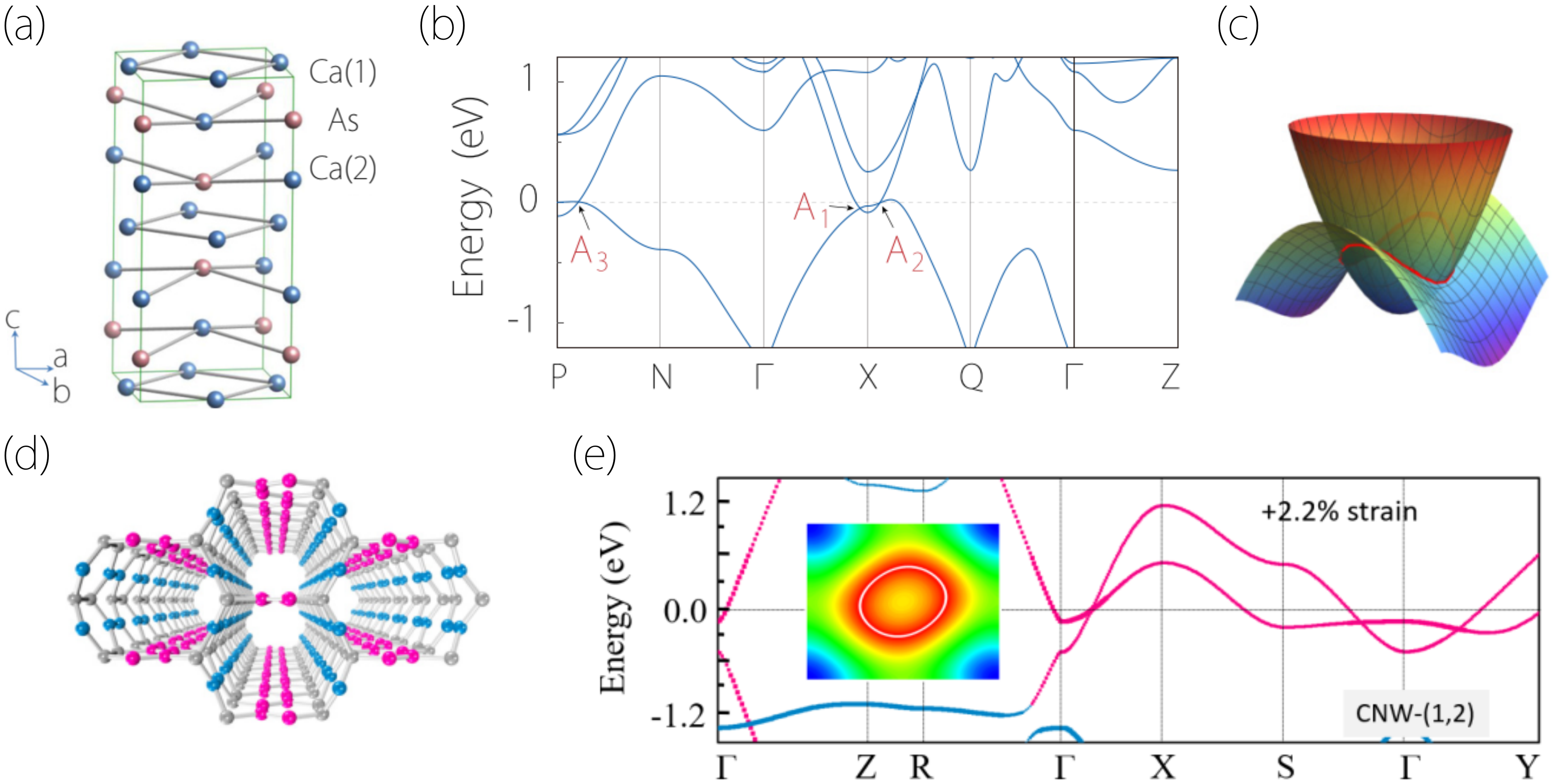}
	\caption{ (a) The crystal structure of Ca$_2$As. (b) Band structure of Ca$_2$As. (c) Band dispersion around $X$ in the $k_z = 0$ plane, showing the hybrid nodal-line.  (d) Crystal structure of 3D carbon allotropes. (e) Under a compressive $+2.2 \%$ strain along $z.$ Inset is a contour plot of the hybrid nodal-line at $k_{z}=0$. Reproduced with permission from Ref.~\cite{zhang2018hybrid,gao2018class}.}
	\label{fig16}
\end{figure}

Several hybrid nodal-line materials have also been proposed, such as the Bernal-stacked graphite~\cite{heikkila2015nexus,hyart2016momentum}, ScCd-type transition-metal intermetallic materials~\cite{li2017type}, Ca$_2$As~\cite{zhang2018hybrid}, certain 3D carbon allotropes~\cite{gao2018class,zhao2019topological}, and B$_2$Si~\cite{li2019new}. The results for Ca$_2$As and a 3D carbon allotrope are shown in Fig.~\ref{fig16}. 

\section{Summary and Outlook}

In this article, we have briefly reviewed the concepts and some recent developments of type-II topological TMs. These materials offer a fertile playground for exploring novel types of emergent fermions. 

Regarding the concept, some studies used the terminology of ``type-III" nodal point. It refers to a special case where one of the crossing bands is flat along the maximally tilted direction. Such a state is not a stable phase, rather, it is a critical point between the type-I and type-II phases, corresponding to the event horizon as discussed in Ref.~\cite{guan2017artificial}.

Looking to the future, there are many interesting questions and directions to be explored. 

First, the study can be extended to other kinds of nodal points, such as the triply degenerate nodal points and the Dirac points without the $\mathcal{PT}$ symmetry. The latter case is also known as the birefringent Dirac point~\cite{kennett2011birefringent,roy2012asymmetric,chen2017ternary}, for which the four crossing bands would generally split along an arbitrary direction. One expects that there could be rich Fermi surface topologies depending on the dispersions around the nodal points.

Second, the classification can also be directly extended to nodal surfaces~\cite{zhong2016towards,liang2016node,turker2018weyl,wu2018nodal}. Depending on the dispersion around each point on the surface, we can naturally have type-I, type-II, and hybrid nodal surfaces, similar to the case for nodal lines. 
In Ref.~\cite{wu2018nodal}, the nodal surfaces are classified based on their protecting mechanisms. The Class-I nodal surfaces have chiral symmetry, so they must belong to type-I. On the other hand, the Class-II surfaces do not have this constraint, so type-II and hybrid types may appear for the Class-II surfaces.

Third, the physical properties of these novel nodal features (including the type-II and hybrid nodal surfaces), especially the interplay between the dispersion and the band topology, still need to be explored. These includes systematic studies on the electric, magnetic, optical, and transport properties. In addition, as discussed in Ref.~\cite{guan2017artificial}, the close analogy with astrophysics can offer intriguing possibility to study astrophysical effects in condensed matter settings.

Last but not least, we need to identify ideal material candidates for these new phases~\cite{tang2019effective}. The desired ones are those with a clean low-energy band structure, with the target nodal feature close to the Fermi level, and with a sufficiently large linear-dispersion region. These requirements are important for experimental studies and for future applications based on these materials.

\acknowledgments{The authors thank D. L. Deng for valuable discussions. {The work is supported by the NSF of China (Grants No.~11734003), the National Key R$\&$D Program of China (Grant No. 2016YFA0300600), the Strategic Priority Research Program of Chinese Academy of Sciences (Grant No. XDB30000000)}, and the Singapore Ministry of Education AcRF Tier 2 (Grant No.~MOE2017-T2-2-108 and MOE2019-T2-1-001).}




%


\end{document}